\documentclass[prd,twocolumn,showpacs,floatfix,amsmath,nofootinbib,amssymb,floatfix]{revtex4}
\usepackage{graphicx,color,dcolumn,booktabs,bm}
\usepackage{longtable,lscape}
\usepackage{txfonts}
\usepackage{overpic}
\usepackage{amssymb}
\usepackage{amsmath}
\usepackage{indentfirst}
\usepackage{feynmf}   
\usepackage{slashed}  
\usepackage{cases}
\usepackage{color}
\usepackage{multirow}
\usepackage{epstopdf}
\usepackage{ulem}
\usepackage{subfigure}
\usepackage[separate-uncertainty]{siunitx}

\newcommand{\vo}{\vec{o}\@ifnextchar{^}{\,}{}}
\def\up{\mathrm}

\graphicspath{{PDF/}}

\def\slash#1{\setbox0=\hbox{$#1$}           
	\dimen0=\wd0                                 
	\setbox1=\hbox{/} \dimen1=\wd1               
	\ifdim\dimen0>\dimen1                        
	\rlap{\hbox to \dimen0{\hfil/\hfil}}      
	#1                                        
	\else                                        
	\rlap{\hbox to \dimen1{\hfil$#1$\hfil}}   
	/                                         
	\fi}                                         %
\def\sl#1{\setbox0=\hbox{#1}
	\dimen0=\wd0
	\rlap{\hbox to \dimen0{\hss/\hss}}%
	#1}

\usepackage[colorlinks, citecolor=blue,anchorcolor=red,menucolor=red, linkcolor=cyan,filecolor=red,runcolor=red,urlcolor=blue,frenchlinks=red]{hyperref}

\def\Tr{\textnormal{Tr}}

 \allowdisplaybreaks

\begin{document}

\title{Study of the hidden charm $D\bar{D}^*$ interactions in chiral effective
	field theory}

\author{Hao Xu$^{1,2}$} \email{xuh2020@nwnu.edu.cn}

\affiliation{
$^1${Institute of Theoretical Physics, College of Physics and Electronic Engineering, Northwest Normal University, Lanzhou 730070, China}\\	
$^2${Lanzhou Center for Theoretical Physics, Lanzhou University, Lanzhou 730000, China}\\
}

\begin{abstract}

{We study the chiral interactions of the hidden charm $D\bar{D}^*$ system within chiral effective field theory. Chiral Lagrangians are constructed by incorporating the chiral symmetry, heavy quark symmetry
as well as proper charge conjugation properties of the heavy mesons. The interacting potentials of the $S$-wave $D\bar{D}^*$ are calculated
up to second chiral order at 1-loop level, where complete two-pion exchange interactions are included. We further investigate the behaviors of the potentials in coordinate space,
 as well as their bound state properties. Our studies indicate that there exists a interacting strength ordering among considered four channels:}   
 $\textnormal{str.}[0^+(1^{++})]>\textnormal{str.}[0^-(1^{+-})] > \textnormal{str.}[1^+(1^{+-})] > \textnormal{str.}[1^-(1^{++})]$ where str. stands for the strength of the $D\bar{D}^*$ interaction.
Moreover, we find that $X(3872)$ can be treated as a good candidate of $0^+(1^{++})$ molecular state. There also tends to form $0^-(1^{+-})$ and $1^+(1^{+-})$ molecular states and we expect
the experiments to search for the predicted multi-structures around the $D\bar{D}^*$ mass region.

\end{abstract}


\maketitle

\section{introduction}\label{sec1}

In past two decades, abundant exotic hadrons have been discovered by upgraded $\tau$-charm 
and $b$ factories such as the BESIII, LHCb, Belle, $BABAR$ and etc. Till now, there has been various forms of exotic quark matters
raising
in hadron spectroscopy: pentaquarks ($P_c$ and $P_{cs}$ states), fully charmed tetraquark candidates (recently discovered $X(6900)$), hidden charm tetraquark candidates (some $XYZ$ states), and etc.
For example, $X(6900)$ was discovered by the LHCb recently \cite{Aaij:2020fnh}, which appears to be a non-trivial
structure in the di-$J/\psi$ invariant mass spectrum. Subsequently, the LHCb Collaboration also reported two structures $X_0(2900)$ and $X_1(2900)$ in the
$B^+\to D^+ D^-K^+$ decay \cite{Aaij:2020ypa}, which are supposed to have four different flavors: $\bar{c}\bar{s}ud$. 
In very recent, the LHCb observed a double charmed structure $T_{cc}$ \cite{LHCb:2021vvq} which is extremely close to $D^0D^{*+}$ threshold 
(the mass difference is $-273\pm61\pm5^{+11}_{-14}$ keV). $T_{cc}$ has a minimal $cc\bar{u}\bar{d}$ content. 
Therefore, one can see it may still be an ongoing progress that other forms of multiquark structures are prepared to be uncovered.    
The rapidly growing numbers of exotic hadrons Urgently demand us to extend our knowledge about non-perturbative QCD.

Although studies of exotic hadrons on the experiment side are in advance, their natures and inner structures are still unclear. 
Theorists try to understand them with all kinds of methods and models, see
Refs.~\cite{Chen:2016qju,Esposito:2016noz,Lebed:2016hpi,Guo:2017jvc,Brambilla:2019esw,Liu:2019zoy} for reviews of theoretical as well as experimental status.

For example, people still can not truly understand the nature of first observed
$XYZ$ state, $X(3872)$ (aka $\chi_{c1}(3872)$). $X(3872)$ was discovered by the Belle collaboration in $B^{+-}\to K^{+-}\pi^+\pi^- J/\psi$ \cite{Choi:2003ue}.
It may be regarded as a charmonium with $\chi_{c1}^\prime(2P)$, but its mass would be much lower than quark model estimate 
(e.g. the GI model calculation in Ref.~\cite{Godfrey:1985xj}). Furthermore, it also has a large decay ratio in the isospin violation process
 $X(3872)\to J/\psi\rho$. It is noteworthy that $X(3872)$ is almost located at $D^0\bar{D}^{*0}$ threshold, so it is believed
that the interaction between $D\bar{D}^{*}$ pair is strongly responsible for the formation of $X(3872)$.

Besides $X(3872)$, there also exists many other $XYZ$ states that may strongly relate to open-charm thresholds.
The charged charmonium-like state $Z_c(3900)$ was observed in the process $e^+e^-\to J/\psi \pi^+\pi^-$ 
\cite{Ablikim:2013mio,Liu:2013dau},
with the mass slightly above $D\bar{D}^*$ threshold it may originate from $D\bar{D}^*$ interaction. Other states such as
$Z_c(4020)$ \cite{Ablikim:2013wzq}, $Y(3940)$ \cite{Abe:2004zs} and $Y(4140)$\cite{Aaltonen:2009tz} are close to respective $D^*\bar{D}^*$
and $D^*_s\bar{D}^*_s$ thresholds too. There also exists some higher $XYZ$ states that are close to excited open-charm  
thresholds: $Z_c^+(4430)$ \cite{Choi:2007wga} with $D^*\bar{D}^{(\prime)}_1$, recently discovered $Y(4626)$ \cite{Jia:2019gfe} (and
$Y(4620)$ \cite{Jia:2020epr}) with $D^*_s \bar{D}_{s1}$, and etc. Taking into account the interactions of these open-charm meson pairs,
people have tried to explain these $XYZ$ states with phenomenological models, such as one boson exchange model \cite{Chen:2016qju,Wang:2021aql}.   

Therefore, for understanding $XYZ$ states mentioned above, it is crucial to elaborately investigate corresponding charmed- 
anticharmed meson interactions within a proper theoretical framework. Chiral effective field theory (ChEFT) is just the powerful formalism that 
satisfies the needs. Other than phenomenological models, in ChEFT, the interactions between hadrons are strictly and
systematically calculated up to a given order, while all contributions such as multi pion exchange are contained in complete. Especially, in our case (i.e. heavy-heavy system) Weinberg's formalism is adopted
\cite{Weinberg:1990rz,Weinberg:1991um}, which has been widely used to investigate the nucleon-nucleon interaction
\cite{Ordonez:1992xp,Ordonez:1995rz,Epelbaum:1998ka,Epelbaum:1999dj,Valderrama:2009ei,Valderrama:2011mv,Long:2011qx,Long:2012ve,Kang:2013uia,Epelbaum:2014efa,Ren:2016jna,Long:2016vnq,Dai:2017ont,Wu:2018lai,Li:2018tbt,Xiao:2018jot,Baru:2019ndr,Ren:2019qow,Xiao:2020ozd,Bai:2020yml,Liu:2020sjo,Wang:2020myr}, see Refs.~\cite{Epelbaum:2008ga,Machleidt:2011zz,Meissner:2015wva,Machleidt:2016rvv,Epelbaum:2019kcf} for reviews. So it is natural to extend the framework to the sector of the heavy hadron interactions.

In Refs.~\cite{Liu:2012vd,Xu:2017tsr}, we have already made attempts to study heavy meson systems with heavy meson chiral effective
field theory (HMChEFT). Ref.~\cite{Liu:2012vd} calculated the potentials of the double bottom system $\bar{B}\bar{B}$ with contact and two pion exchange mechanisms under
HMChEFT. Ref.~\cite{Xu:2017tsr} developed corresponding techniques and investigated the double charmed system $D D^*$, then further utilized
the Schr\"odinger equation to search for bound state solutions. Following the same procedures,
authors in Refs.~\cite{Wang:2018atz,Meng:2019ilv,Wang:2019nvm,Wang:2019ato,Meng:2019nzy,Wang:2020dhf,Chen:2021htr} studied other heavy hadron systems.

With the experience in the studies of double heavy-flavored systems above, it is natural to extend to the heavy-antiheavy 
flavored systems, which will directly link to the charmonium-like states mentioned before. 
Note that this extension is not straightforward, for example, the interactions in the heavy-antiheavy 
flavored systems should be 
constructed by properly considering charge conjugation properties
of the fields in the heavy quark limit.

In this work, choosing the 
charmed-anticharmed system $D\bar{D}^*$ as an example, we will try to study $D\bar{D}^*$ interactions up to second chiral order $O(\epsilon^2)$
at 1-loop level using weinberg's scheme. The contributions of contact, one pion exchange (OPE) and two pion
exchange (TPE)
will be included in complete. As mentioned above, different from calculated double heavy-flavored systems before \cite{Liu:2012vd,Xu:2017tsr}, 
additional symmetries (the charge conjugation symmetry) as well as proper charge conjugation states should be concerned when constructing $D\bar{D}^*$ interactions and
$D\bar{D}^*$ scattering amplitudes.

Following the strategy in Ref.~\cite{Xu:2017tsr}, we will iterate obtained $D\bar{D}^*$ potentials into
the Schr\"odinger equation, to see whether the $D\bar{D}^*$ interactions are strong enough to form bound states. It is noteworthy that
under the one-boson-exchange model, Refs.~\cite{Liu:2008fh,Liu:2008tn,Li:2012cs,Zhao:2014gqa,Wang:2021aql} also studied the $D\bar{D}^*$ system,
they considered one boson ($\pi$, $\sigma$, $\rho$, $\omega$ and etc.) exchange mechanism.
They fond that, $D\bar{D}^*$ in $J^{PC}=1^{++}$, $I=0$ channel ($X(3872)$'s $J^{PC}$), as well as some other channels,
are strong enough to form a bound state.
Therefore, results and conclusions presented in this work may be a comparison to theirs. Also, possible $D\bar{D}^*$ molecular states have been studied extensively in various methods
\cite{Wong:2003xk,Swanson:2003tb,Suzuki:2005ha,Thomas:2008ja,Lee:2009hy,He:2014nya,
Voloshin:2003nt,Close:2003sg,Tornqvist:2004qy,Sun:2011uh,Sun:2012zzd,Wang:2017dcq,Ding:2020dio,Sun:2017wgf,Yang:2017prf,Ding:2009vj,Zhang:2006ix,Tornqvist:1993ng,DeRujula:1976zlg}.

This paper is organized as follows. In Sec.~\ref{SecLagrangian} we describe concerned $D\bar{D}^*$ Lagrangians by considering the chiral symmetry and heavy quark symmetry,
as well as by properly taking into account charge conjugation properties of the heavy mesons. In Sec.~\ref{SecPotential} we calculate the potentials of $D\bar{D}^*$ up to second chiral order $O(\epsilon^2)$
at 1-loop level using weinberg's scheme. In Sec.~\ref{SecN}, after solving the Schr\"odinger equation with calculated $D\bar{D}^*$ potentials, we investigate the bound state properties in
considered four channels. Then we discuss the behaviors of the potentials in coordinate space to further understand $D\bar{D}^*$ interactions and the mechanisms of the bound state formations. Later we discuss obtained molecular states and 
their discovery potantials in detail. The last section denotes to a summary.

\section{Chiral Lagrangians of $D\bar{D}^*$ system in HMChEFT} \label{SecLagrangian}
Like Refs.~\cite{Liu:2012vd,Xu:2017tsr}, we adopt HMChEFT,  and derive the Lagrangians and effective potentials in a strict 
power-counting scheme. In this framework, the amplitudes or potentials are arranged according to the chiral order 
$\epsilon=p/\Lambda_\chi$ ($p$ stands for momentum of a pion, residual momentum of a heavy meson, or $D$-$D^*$ mass
splitting). 
In this work flavor $SU(2)$ symmetry is considered.

We first show the Lagrangians of concerned $D\bar{D}^*$ system at leading order. First, the $DD^*\pi$ Lagrangian at $O(\epsilon^1)$ is
needed \cite{Burdman:1992gh,Wise:1992hn,Yan:1992gz}:
\begin{eqnarray}\label{LagrangianHpi1}
\mathcal L^{(1)}_{H\phi}&=&-\langle (i v\cdot \partial H)\bar H
\rangle
+\langle H v\cdot \Gamma \bar H \rangle
+g\langle H \slashed u \gamma_5 \bar H\rangle
\nonumber \\  &&-\frac18 \delta \langle H \sigma^{\mu\nu} \bar H \sigma_{\mu\nu} \rangle,
\end{eqnarray}
where $H$ field describing the $(D,D^*)$ doublet is
\begin{eqnarray} \label{Hfield}
&& H=\frac{1+\slashed v}{2}\left(P^*_\mu\gamma^\mu+iP\gamma_5\right),\quad \nonumber \\
&&\bar H=\gamma^0 H^\dag \gamma^0 = \left(P^{*\dag}_\mu \gamma^\mu+iP^\dag \gamma_5\right) \frac{1+\slashed v}{2},\nonumber\\
&& P=(D^0, D^+), \quad P^*_\mu=(D^{*0}, D^{*+})_\mu.
\end{eqnarray}
$v=(1,0,0,0)$ stands for the 4-velocity of the $H$ field. The axial vector field $u$ and chiral connection $\Gamma$ are expressed as
\begin{equation}
\Gamma_\mu = {i\over 2} [\xi^\dagger, \partial_\mu\xi],\quad
u_\mu={i\over 2} \{\xi^\dagger, \partial_\mu \xi\},
\end{equation}
where $\xi =\exp(i \phi/2f)$, $f$ is the bare pion constant, and
\begin{equation}
\phi=\sqrt2\left(
\begin{array}{cc}
\frac{\pi^0}{\sqrt2}&\pi^+\\
\pi^-&-\frac{\pi^0}{\sqrt2}\\
\end{array}\right).
\end{equation}

For studying the $D\bar{D}^*$ system under HMChEFT, we also need to describe the interaction between an anti-charmed meson and a pion. 
Applying charge conjugation transformation to Eq.~(\ref{LagrangianHpi1}), the Lagrangian of the interacting part is given by
\begin{eqnarray}\label{LagrangianHcpi1}
\mathcal L^{(1)}_{H_c\phi}&=&
+\langle \bar{H}_c v\cdot \Gamma  H_c \rangle +g\langle \bar{H}_c \slashed u \gamma_5 H_c\rangle,
\end{eqnarray}
where $g=0.59$. In the above, $H_c$ field represents the anti-charmed meson doublet $(\bar{D},\bar{D}^*)$ in the heavy quark limit, 
and subscript $c$ stands for charge conjugation. Note that $H_c$ is defined as charge conjugation of $H$ in Eq.~(\ref{LagrangianHpi1}):
\begin{eqnarray} \label{Hcfield}
&& H_c=\left(P^*_{c\mu}\gamma^\mu+iP_c\gamma_5\right)\frac{1-\slashed v}{2},\quad \nonumber \\
&&\bar H_c=\gamma^0 H^\dag_c \gamma^0 = \frac{1-\slashed v}{2}\left(P^{*\dag}_{c\mu} \gamma^\mu+iP^\dag_c \gamma_5\right) ,\nonumber\\
&& P_c=(\bar{D}^0, D^-, D^-_s), \quad P^*_{c\mu}=(\bar{D}^{*0}, D^{*-},D^{*-}_s)_\mu.
\end{eqnarray}

Then, the contact Lagrangian at $O(\epsilon^0)$ is needed for $O(\epsilon^0)$ and $O(\epsilon^2)$ contact amplitudes,
which can be constructed as
  \begin{eqnarray}\label{Lagrangian4H0}
\mathcal L^{(0)}_{2H2H_c}&=&D_{a} \textnormal{Tr} [H \gamma_\mu \bar H ] \textnormal{Tr} [\bar H_c
\gamma^\mu H_c] \nonumber \\  &&+D_{b} \textnormal{Tr} [H \gamma_\mu\gamma_5 \bar H ] \textnormal{Tr} [\bar H_c
\gamma^\mu\gamma_5 H_c] \nonumber\\  && +E_{a} \textnormal{Tr} [H
\gamma_\mu\tau^a \bar H ] \textnormal{Tr} [\bar H_c \gamma^\mu\tau_a H_c] \nonumber\\
&&+E_{b} \textnormal{Tr} [H \gamma_\mu\gamma_5\tau^a \bar H ] \textnormal{Tr} [\bar H_c
\gamma^\mu\gamma_5\tau_a H_c],
\end{eqnarray}
where $D_a$, $D_b$, $E_a$, $E_b$ are four independent low energy constants (LECs).

For the loop diagrams in the potentials (or the amplitudes) at order $O(\epsilon^2)$, we also need $O(\epsilon^2)$ contact 
Lagrangians to cancel their divergences:
\begin{align}
&\mathcal L^{(2,h)}_{2H2H_c} \nonumber\\
&= D_{a}^h\Tr[H \gamma_\mu \bar H ]\Tr[ \bar H_c
\gamma^\mu H_c] \Tr(\chi_+) \nonumber \\ 
&\quad \! +D_{b}^h\Tr[H \gamma_\mu\gamma_5
\bar H ]\Tr[\bar H_c \gamma^\mu\gamma_5 H_c]\Tr(\chi_+) \nonumber\\
&\quad \!+E_{a}^h \Tr[H \gamma_\mu\tau^a \bar H ]\Tr[\bar H_c
\gamma^\mu\tau_a H_c]\Tr(\chi_+)  \nonumber \\ 
&\quad \!+E_{b}^h \Tr[H\gamma_\mu\gamma_5\tau^a \bar H ]\Tr[\bar H_c
\gamma^\mu\gamma_5\tau_a H_c]\Tr(\chi_+),  \label{Lagrangian4H2h}
\\
&\mathcal L^{(2,v)}_{2H2H_c}\nonumber\\ 
&= D_{a1}^v \bigg( \Tr[(v\cdot D H) \gamma_\mu
(v\cdot D \bar H) ]\Tr[\bar H_c \gamma^\mu H_c] + \text{C.c.} \bigg  ) \nonumber \\ 
& \quad \! +D_{a2}^v \bigg( \Tr[(v\cdot D
H) \gamma_\mu \bar H ]\Tr[ (v\cdot D \bar H_c) \gamma^\mu H_c] +\text{C.c.} \bigg)
\nonumber\\
& \quad \! +D_{a3}^v \bigg( \Tr[(v\cdot D H) \gamma_\mu \bar H ]\Tr[\bar  H_c
\gamma^\mu(v\cdot D  H_c)] +\text{H.c.} \bigg) \nonumber \\
 & \quad \! +D_{a4}^v \bigg( \Big[ \Tr[((v\cdot D)^2 H)
\gamma_\mu \bar H ]\Tr[ \bar H_c \gamma^\mu H_c ] +\text{H.c.} \Big]+\text{C.c.} \bigg) \nonumber\\
& \quad \! +D_{b1}^v \bigg(\Tr[(v\cdot D H) \gamma_\mu\gamma_5 (v\cdot D \bar H)
]\Tr[\bar H_c \gamma^\mu\gamma_5 H_c] +\text{C.c.} \bigg) \nonumber\\
& \quad \! +... \nonumber \\ 
& \quad \! +E_{a1}^v \bigg( \Tr[(v\cdot D H)
\gamma_\mu\tau^a (v\cdot D \bar H) ]\Tr[\bar  H_c
\gamma^\mu\tau_aH_c]+ \text{C.c} \bigg) \nonumber\\
&  \quad \! +... \nonumber\\
& \quad \! +E_{b1}^v \bigg( \Tr[(v\cdot
D H) \gamma_\mu\gamma_5\tau^a (v\cdot D \bar H) ]\Tr[\bar H_c
\gamma^\mu\gamma_5\tau_a  H_c] +\text{C.c.}  \bigg) \nonumber\\ 
&  \quad \!+... ,\label{Lagrangian4H2v}
\end{align}

\begin{align}
&\mathcal L^{(2,q)}_{2H2H_c} \nonumber\\
&= D_1^q \bigg( \Big[ \Tr[(D^\mu H) \gamma_\mu\gamma_5
(D^\nu \bar H) ]\Tr[ \bar{H}_c \gamma_\nu\gamma_5 H_c]  +\text{H.c.} \Big]  +\text{C.c.}  \bigg) \nonumber \\
&  \quad \!  +D_2^q \bigg(  \Tr[(D^\mu
H) \gamma_\mu\gamma_5 \bar H ]\Tr[ (D^\nu \bar{H}_c)
\gamma_\nu\gamma_5 H_c]   +\text{C.c.} \bigg) \nonumber\\
& \quad \!  +D_3^q \bigg( \Tr[(D^\mu H)
\gamma_\mu\gamma_5 \bar H ]\Tr[ \bar{H}_c \gamma_\nu\gamma_5(D^\nu 
H_c)] +\text{H.c.} \bigg) \nonumber \\ 
& \quad \!  +D_4^q \bigg( \Big[\Tr[(D^\mu D^\nu H) \gamma_\mu\gamma_5 \bar H ]\Tr[ \bar{H}_c
\gamma_\nu\gamma_5 H_c] +\text{H.c.} \Big]  +\text{C.c.}  \bigg) \nonumber\\
& \quad \!  +E_1^q \bigg( \Big[ \Tr[(D^\mu H)
\gamma_\mu\gamma_5 \tau^a(D^\nu \bar H) ]\Tr[ \bar{H}_c
\gamma_\nu\gamma_5\tau_a H_c]+\text{H.c.} \Big] \nonumber\\
& \quad \! \quad \! +\text{C.c.}  \bigg) \nonumber \\ 
&  \quad \!  +...,  \label{Lagrangian4H24}
\end{align} 
where
\begin{eqnarray}
\chi_\pm&=&\xi^\dagger\chi\xi^\dagger\pm\xi\chi\xi , \nonumber\\
\chi&=&m_\pi^2.
\end{eqnarray}
In above Lagrangians, H.c. and C.c. stand for Hermitian conjugation and charge conjugation respectively.  Notice that finite parts of above
Lagrangians could also contribute to the potentials, however they will bring a large number
of LECs.

In present work, we will adopt Weinberg's formalism \cite{Weinberg:1990rz,Weinberg:1991um}. Previous works
\cite{Liu:2012vd,Xu:2017tsr} have already applied Weinberg's power counting scheme to investigate $\bar{B}\bar{B}$ and $DD^*$
systems. In a word, the formalism states that with standard power counting scheme, one first calculates effective potentials 
(the sum of the two-particle irreducible (2PI) diagrams), then uses them to solve the Lippmann-Schwinger or Schrödinger equation, 
so complete contributions including two-particle reducible (2PR) diagrams can be retrieved. Here we refer to Refs.~\cite{Liu:2012vd,Xu:2017tsr} for more details. 

\section{potentials of the $D\bar{D}^*$ system in HMChEFT} \label{SecPotential}

We will investigate four different channels in the $D\bar{D}^*$ system: $C$ parity $C=\pm1$ and isospin $I=0$, $1$. Therefore we
consider following flavor wave functions: 
\begin{align} \label{DDstarbFlavorWF}
&\Big|0,0 \Big\rangle=\frac12\bigg[ \Big( \big|D^0\bar{D}^{*0}\big\rangle+\big|D^+D^{*-}\big\rangle \Big)
+c \Big( \big|D^{*0}\bar{D}^{0}\big\rangle+\big|D^{*+}D^{-}\big\rangle \Big)\bigg] \ , \nonumber\\
&\Big|1,0 \Big\rangle=\frac12\bigg[ \Big( \big|D^0\bar{D}^{*0}\big\rangle-\big|D^+D^{*-}\big\rangle \Big)
+c \Big( \big|D^{*0}\bar{D}^{0}\big\rangle-\big|D^{*+}D^{-}\big\rangle \Big)\bigg] \ ,
\nonumber\\
&\Big|1,1 \Big\rangle=\frac{1}{\sqrt{2}} \Big( \big|D^+\bar{D}^{*0}\big\rangle+ c\big|D^{*+}\bar{D}^{0}\big\rangle \Big)\ ,
\nonumber\\
&\Big|1,-1 \Big\rangle=\frac{1}{\sqrt{2}} \Big( \big|D^{*0}D^-\big\rangle+ c\big|D^{0}D^{*-}\big\rangle \Big)\ ,
\end{align}
where parameter $c=\mp$ stands for $C=\pm1$.   

We first calculate the elastic scattering amplitudes in processes $D\bar{D}^*\to D\bar{D}^*$ with defined $D\bar{D}^*$ states (\ref{DDstarbFlavorWF}). 
The contributions up to
$O(\epsilon^2)$ are considered. Accoridng to the chiral Lagrangians in previous, there exists tree-level contact and OPE diagrams
at lowest order $O(\epsilon^0)$. While at $O(\epsilon^2)$, there emerges 1-loop contact and OPE diagrams, as 
well as TPE contributions.

\begin{figure}[htpb]
	\begin{center}
		\includegraphics[scale=0.45]{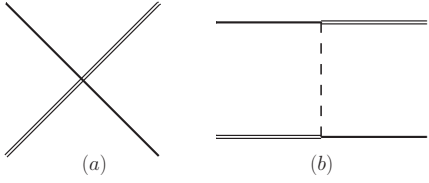}
		\caption{Tree-level diagrams of the $D\bar{D}^{*}$ system
			at $O(\epsilon^0)$. The solid, double-solid, and dashed lines stand for $D$ (or $\bar{D}$), $D^{*}$ (or $\bar{D}^{*}$), and $\pi$, respectively.}\label{O0TreeDiagram}
	\end{center}
\end{figure}

Let us focus on the order $O(\epsilon^0)$ first. At $O(\epsilon^0)$, there are two tree-level diagrams which are depicted in Fig.~\ref{O0TreeDiagram}. Obviously, Fig.~\ref{O0TreeDiagram}(a) stands for contact contribution while
Fig.~\ref{O0TreeDiagram}(b) stands for OPE contribution. 

Here we stress that in our depicted diagrams
(such as in Fig.\ref{O0TreeDiagram}), (double-) solid line stands for $D^{(*)}$ as well as anti-particle $\bar{D}^{(*)}$, 
depending on the concrete term
in the expanded isospin amplitude $\langle I,0|T|I,0\rangle$, where $|I,0\rangle$ is defined in Eq.~(\ref{DDstarbFlavorWF}).
Notice that we label the momenta of external fields as: $p_1$ for initial $D$, $p_2$ for initial $\bar{D}^*$, $p_3$ for final $D$
and $\bar{D}$, $p_4$ for final $D^*$ and $\bar{D}^*$

At this order, the vertexes are also at $O(\epsilon^0)$, therefore we use following $O(\epsilon^0)$ Lagrangians:
the Lagrangian (\ref{LagrangianHpi1}) which depicts
the $DD^*\pi$ vertex, the Lagrangian (\ref{LagrangianHcpi1}) describing $\bar{D}\bar{D}^*\pi$ vertex, and the contact
$DD^*\bar{D}\bar{D}^*$ Lagrangian (\ref{Lagrangian4H0}). Consequently, the contact amplitudes for the diagram of
Fig.~\ref{O0TreeDiagram}(a) are calculated to be
\begin{align}\label{AmplitudeContact0}
&\mathcal{M}^{(0)}_{(a)} = 4 (D_a+3E_a-cD_b-3cE_b)\varepsilon(p_2) \cdot \varepsilon^*(p_4)  \quad \textnormal{ for $I=0$},
\\
&\mathcal{M}^{(0)}_{(a)} = 4 (D_a-E_a-cD_b+cE_b)\varepsilon(p_2) \cdot \varepsilon^*(p_4)   \qquad \textnormal{ for $I=1$}.
\end{align}
The OPE contributions of Fig.~\ref{O0TreeDiagram}(b) read
\begin{align}\label{AmplitudeOPE00}
&\mathcal{M}^{(0)}_{(b)} = 3c \frac{g^2}{f^2} \frac{1}{p^2-m^2} p\cdot\varepsilon(p_2) p\cdot\varepsilon^*(p_4) \quad \textnormal{ for $I=0$},
\\
&\mathcal{M}^{(0)}_{(b)} = -c \frac{g^2}{f^2} \frac{1}{p^2-m^2} p\cdot\varepsilon(p_2) p\cdot\varepsilon^*(p_4) \quad \textnormal{ for $I=1$}.
\label{AmplitudeOPE01}
\end{align}
In the above, $p=p_1-p_4$ denotes the momentum transfer, the superscript (0) of $\mathcal{M}$ stands for the chiral
order $O(\epsilon^0)$. The parameter $c$ appearing in $\mathcal{M}$ has been defined in Eq.~(\ref{DDstarbFlavorWF}), which 
takes $\mp1$ for $C=\pm1$.

\begin{figure}[htpb]
	\begin{center}
		\includegraphics[scale=0.38]{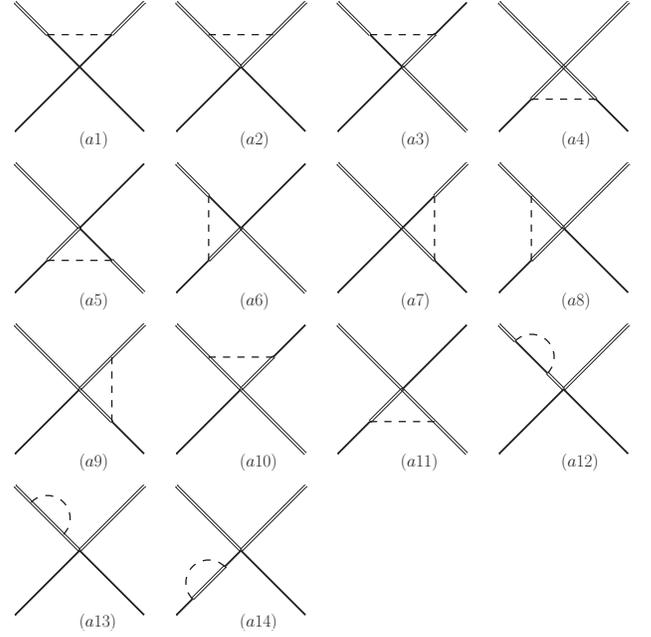}
		\caption{Contact diagrams at $O(\epsilon^2)$. The solid, double-solid, and dashed lines stand for $D$ (or $\bar{D}$), $D^{*}$ (or $\bar{D}^{*}$) and $\pi$, respectively.}\label{O2ContactDiagram}
	\end{center}
\end{figure}

\begin{figure}[htpb]
	\begin{center}
		\includegraphics[scale=0.43]{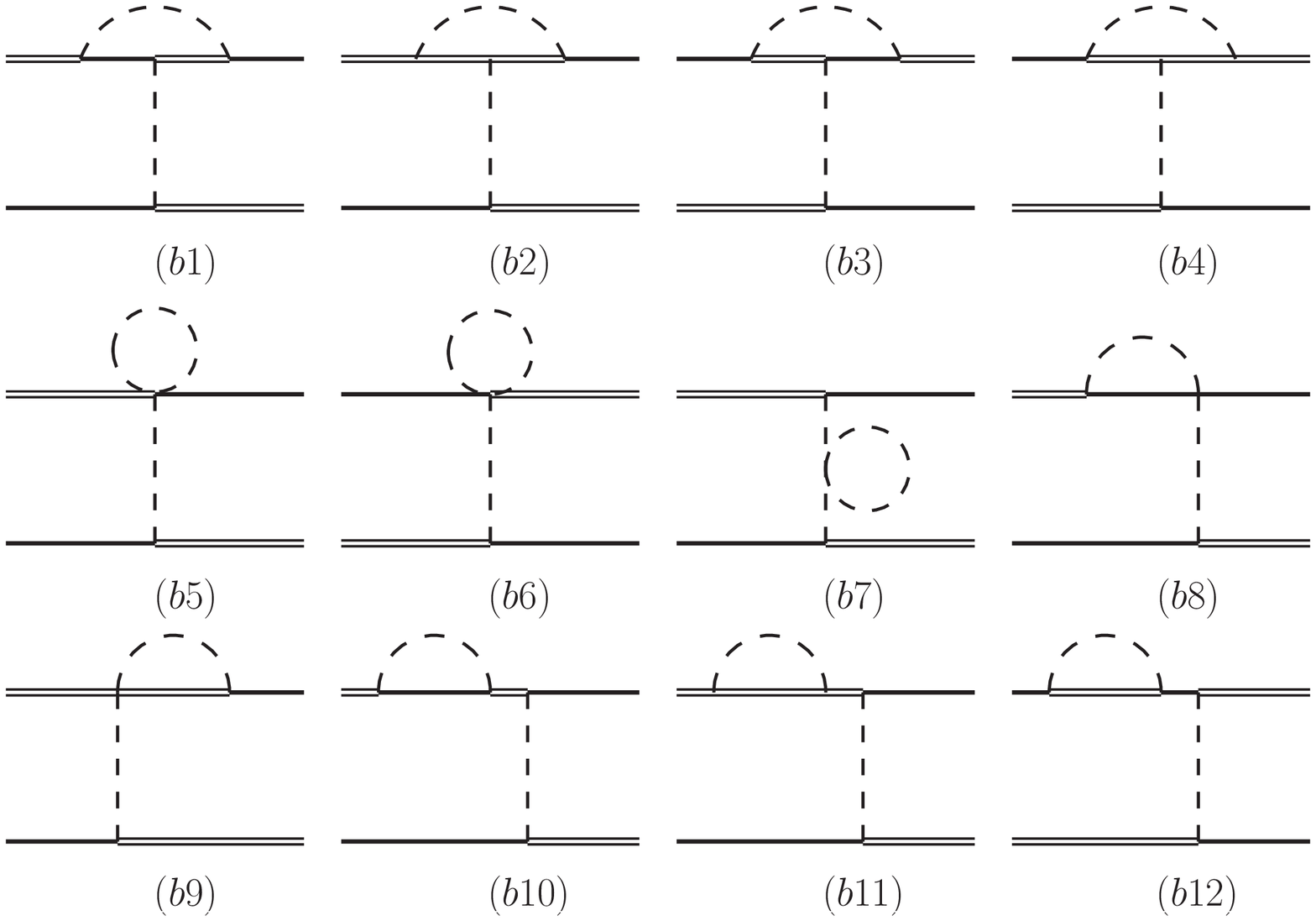}
		\caption{ OPE diagrams at $O(\epsilon^2)$. The solid, double-solid, and dashed lines stand for $D$ (or $\bar{D}$), $D^{*}$ (or $\bar{D}^{*}$) and $\pi$, respectively.}\label{O21piDiagram}
	\end{center}
\end{figure}

\begin{figure}[htpb]
	\begin{center}
		\includegraphics[scale=0.35]{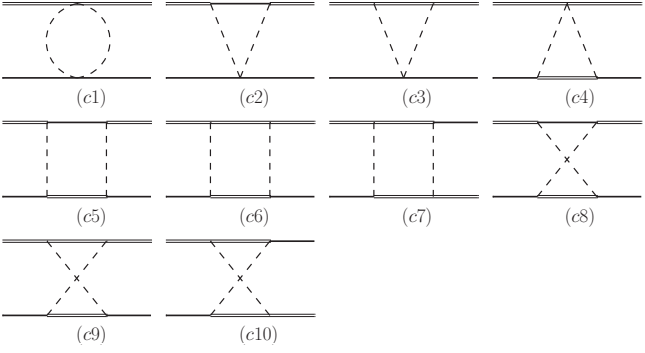}
		\caption{TPE diagrams at $O(\epsilon^2)$. The solid, double-solid, and dashed lines stand for $D$ (or $\bar{D}$), $D^{*}$ (or $\bar{D}^{*}$) and $\pi$, respectively.}\label{O22piDiagram}
	\end{center}
\end{figure}

Then, we consider the contributions at order $O(\epsilon^2)$, which are illustrated in Figs.~\ref{O2ContactDiagram}-\ref{O22piDiagram}.
We can see that there are three types of the diagrams. The diagrams in Fig.~\ref{O2ContactDiagram} stand for $O(\epsilon^2)$ contact
interactions, which are 1-loop corrections to Fig.~\ref{O0TreeDiagram}(a). In Fig.~\ref{O21piDiagram}, 
the diagrams all contribute to OPE interactions, which are just 1-loop corrections to Fig.~\ref{O0TreeDiagram}(b).
Besides contact and OPE contrbutions, there appears new types of diagrams at this order: TPE interactions. They are
showing in Fig.~\ref{O22piDiagram}. 

Notice that Figs.~\ref{O2ContactDiagram}-\ref{O22piDiagram} may only represent typical diagrams having different topologies.
For example, each contact diagram showing in Fig.~\ref{O2ContactDiagram} represents both direct channel
$D\bar{D}^{*}\to D\bar{D}^{*}$ and cross channel $D\bar{D}^{*}\to \bar{D}D^{*}$.
The combination of these two channels depends on the specific isospin amplitude $\langle I,0|T|I,0\rangle$, where
direct channel and cross channel generally both appear. 
In formal, those typical diagrams in Figs.~\ref{O2ContactDiagram}-\ref{O22piDiagram} are
the same as the double charmed system $DD^*$'s investigated in our previous work \cite{Xu:2017tsr}, i.e., these two systems
has the same topology in Feynmann diagrams. Although, there is no cross channel in $DD^*$ system. 

We now focus on the $O(\epsilon^2)$ contact contributions first (Fig.~\ref{O2ContactDiagram}). Involved vertexes are
$O(\epsilon^0)$ contact interaction (\ref{Lagrangian4H0}), $O(\epsilon^0)$ $DD^*\pi$ interaction (\ref{LagrangianHpi1}) and
$O(\epsilon^0)$ $\bar{D}\bar{D}^*\pi$ interaction (\ref{LagrangianHcpi1}). Combined with defined flavor wave
function (\ref{DDstarbFlavorWF}), their isospin amplitudes can be written as
\begin{align}\label{AmplitudeContact2I}
&\mathcal{M}^{(2)}_{(a1)}= -4\frac{g^2}{f^2} g_{s1} (A_0-cA_1) J^g_{22} \varepsilon \cdot \varepsilon^*,
\\
&\mathcal{M}^{(2)}_{(a2)}= -4(-3+d)(-2+d)\frac{g^2}{f^2} (A_0 g_{s1}-c A_1 g_{s5}) J^g_{22} \varepsilon \cdot \varepsilon^*,
\\
&\mathcal{M}^{(2)}_{(a3)}= -4\frac{g^2}{f^2} (A_0-cA_1) g_{s5} J^g_{22} \varepsilon \cdot \varepsilon^*,
\\
&\mathcal{M}^{(2)}_{(a4)}= -4\frac{g^2}{f^2} \Big( -A_0 g_{s1}+d A_0 g_{s1}-c A_1 g_{s1}+2c A_1 g_{s5}
\nonumber\\
&\qquad\quad\ -c d A_1 g_{s5} \Big) J^g_{22} \varepsilon \cdot \varepsilon^*,
\\
&\mathcal{M}^{(2)}_{(a5)}= -4\frac{g^2}{f^2} (A_0-cA_1)g_{s5} J^g_{22} \varepsilon \cdot \varepsilon^*,
\\
&\mathcal{M}^{(2)}_{(a6)}= -4\frac{g^2}{f^2} (A_0 g_{s5}-cA_1 g_{s1}) J^h_{22} \varepsilon \cdot \varepsilon^*,
\\
&\mathcal{M}^{(2)}_{(a7)}= -4\frac{g^2}{f^2} (A_0 g_{s5}-cA_1 g_{s1}) J^h_{22} \varepsilon \cdot \varepsilon^*,
\\
&\mathcal{M}^{(2)}_{(a8)}= -4(d-3)(d-2)\frac{g^2}{f^2} (A_0+cA_1)g_{s5} J^h_{22} \varepsilon \cdot \varepsilon^*,
\\
&\mathcal{M}^{(2)}_{(a9)}= -4(d-3)(d-2)\frac{g^2}{f^2} (A_0+cA_1)g_{s5} J^h_{22} \varepsilon \cdot \varepsilon^*,
\\
&\mathcal{M}^{(2)}_{(a10)}= -4(d-3)(d-2)\frac{g^2}{f^2} (A_0+cA_1)g_{s5} J^g_{22} \varepsilon \cdot \varepsilon^*,
\\
&\mathcal{M}^{(2)}_{(a11)}= -4(d-3)(d-2)\frac{g^2}{f^2} (A_0+cA_1)g_{s5} J^g_{22} \varepsilon \cdot \varepsilon^*,
\\
&\mathcal{M}^{(2)}_{(a12+13)}= -\frac32 \frac{g^2}{f^2} (A_0 g_{s1}-cA_1 g_{s5}) \Big[(d-2)\partial_\omega J^b_{22}(\omega_1)
\nonumber\\
&\qquad\qquad \quad  +\partial_\omega J^b_{22}(\omega_2)\Big]\varepsilon \cdot \varepsilon^*,
\\
&\mathcal{M}^{(2)}_{(a14)}= -\frac32 (d-1)\frac{g^2}{f^2} (A_0 g_{s1}-cA_1 g_{s5}) \partial_\omega J^b_{22}\varepsilon \cdot \varepsilon^*.
\label{AmplitudeContact2F}
\end{align}

\renewcommand{\arraystretch}{1.5}
\begin{table*}[!htbp]
	\centering
	\caption{The constants appearing in the contact amplitudes (Eqs.~(\ref{AmplitudeContact2I})-(\ref{AmplitudeContact2F})). 
	}\label{TabContactA}
	\begin{tabular}{c|cc|cc|cc}	\toprule[1pt]
		& \multicolumn{2}{c|}{ $I=0$  } &\multicolumn{2}{c|}{ $I=1$ } &       & \\
	               &  $A_0$   &    $A_1$  &  $A_0$   &    $A_1$   &    $\omega_1$   & $\omega_2$ \\
		\midrule[1pt]
$A_{a1}$ & $\frac34 (g_{f1}-g_{f\lambda})$ & $\frac34 (g_{f1}-g_{f\lambda})$ & $\frac34 g_{f1}+\frac14 g_{f\lambda}$ & $-\frac14 g_{f1}+\frac54 g_{f\lambda}$ & $\delta$ & $\delta$ \\
$A_{a2}$ & $\frac34 (g_{f1}-g_{f\lambda})$ & $\frac34 (g_{f1}-g_{f\lambda})$ & $\frac34 g_{f1}+\frac14 g_{f\lambda}$ & $-\frac14 g_{f1}+\frac54 g_{f\lambda}$ &    $0$   &  $0$ \\
$A_{a3}$ & $\frac34 (g_{f1}-g_{f\lambda})$ & $\frac34 (g_{f1}-g_{f\lambda})$ & $-\frac14 g_{f1}+\frac54 g_{f\lambda}$ & $\frac34 g_{f1}+\frac14 g_{f\lambda}$ &  $\delta$ & $-\delta$ \\
$A_{a4}$ & $\frac34 (g_{f1}-g_{f\lambda})$ & $\frac34 (g_{f1}-g_{f\lambda})$ & $\frac34 g_{f1}+\frac14 g_{f\lambda}$ & $-\frac14 g_{f1}+\frac54 g_{f\lambda}$ & $-\delta$ & $-\delta$ \\
$A_{a5}$ & $\frac34 (g_{f1}-g_{f\lambda})$ & $\frac34 (g_{f1}-g_{f\lambda})$ & $-\frac14 g_{f1}+\frac54 g_{f\lambda}$ & $\frac34 g_{f1}+\frac14 g_{f\lambda}$ & $-\delta$ & $\delta$ \\
$A_{a6}$ & $\frac34 g_{f1}+\frac94 g_{f\lambda}$ & $\frac34 g_{f1}+\frac94 g_{f\lambda}$ & $\frac14(-g_{f1}+g_{f\lambda})$ & $\frac14(-g_{f1}+g_{f\lambda})$ & $\delta$ & $-\delta$ \\
$A_{a7}$ & $\frac34 g_{f1}+\frac94 g_{f\lambda}$ & $\frac34 g_{f1}+\frac94 g_{f\lambda}$ & $\frac14(-g_{f1}+g_{f\lambda})$ & $\frac14(-g_{f1}+g_{f\lambda})$ & $-\delta$ & $\delta$\\
$A_{a8}$ & $\frac34 g_{f1}+\frac94 g_{f\lambda}$ & $\frac34 g_{f1}+\frac94 g_{f\lambda}$ & $\frac14(-g_{f1}+g_{f\lambda})$ & $\frac14(-g_{f1}+g_{f\lambda})$ & $0$ & $-\delta$ \\
$A_{a9}$ & $\frac34 g_{f1}+\frac94 g_{f\lambda}$ & $\frac34 g_{f1}+\frac94 g_{f\lambda}$ & $\frac14(-g_{f1}+g_{f\lambda})$ & $\frac14(-g_{f1}+g_{f\lambda})$ & $-\delta$ & $0$ \\
$A_{a10}$ & $\frac34 (g_{f1}-g_{f\lambda})$ & $\frac34 (g_{f1}-g_{f\lambda})$ & $-\frac14 g_{f1}+\frac54 g_{f\lambda}$ & $\frac34 g_{f1}+\frac14 g_{f\lambda}$ & $0$ & $-\delta$ \\
$A_{a11}$ & $\frac34 (g_{f1}-g_{f\lambda})$ & $\frac34 (g_{f1}-g_{f\lambda})$ & $-\frac14 g_{f1}+\frac54 g_{f\lambda}$ & $\frac34 g_{f1}+\frac14 g_{f\lambda}$ & $-\delta$ &  $0$  \\
$A_{a12+13}$ &   $g_{f1}+3g_{f\lambda}$  &   $g_{f1}+3g_{f\lambda}$  & $g_{f1}-g_{f\lambda}$ & $g_{f1}-g_{f\lambda}$ & $0$ & $\delta$ \\
$A_{a14}$ &   $g_{f1}+3g_{f\lambda}$  &   $g_{f1}+3g_{f\lambda}$  & $g_{f1}-g_{f\lambda}$ & $g_{f1}-g_{f\lambda}$ & $-\delta$ & $-$ \\
		\bottomrule[1pt]
	\end{tabular}
\end{table*}

In the above, $c$ still takes $\mp$ for $C=\pm$. $d$ is the space-time dimension coming from the dimensional regularization. $A_0$ and $A_1$ are
constants depending on different diagrams and isospin $I$, which are collected in Table \ref{TabContactA}. Also, in these expressions
$\varepsilon$ and $\varepsilon^*$ are the abbreviations of the polarization vectors $\varepsilon(p_2)$ and $\varepsilon^*(p_4)$ respectively.
Each $J$ is a loop function defined in Refs.~\cite{Liu:2012vd,Xu:2017tsr}. For following OPE and TPE amplitudes these notations apply.

Notice that, one should further expand these amplitudes (\ref{AmplitudeContact2I})-(\ref{AmplitudeContact2F}), later variables $g_{si}$ and $g_{fi}$
(embed in the constants $A_i$) in them have to be replaced as the following: 
\begin{align}
&g_{s1}g_{f1} \to D_a, \qquad g_{s5}g_{f1} \to D_b,  \nonumber\\
& g_{s1}g_{f\lambda} \to E_a, \qquad g_{s5}g_{f\lambda} \to E_b.
\end{align}

Then we consider the OPE contributions showing in Fig.~\ref{O21piDiagram}. To describe $O(\epsilon^0)$ $DD^*\pi$, $DD^*3\pi$, $\bar{D}\bar{D}^*\pi$ 
and $\bar{D}\bar{D}^*3\pi$ vertexes, we utilize the chiral Lagrangians (\ref{LagrangianHpi1}) and (\ref{LagrangianHcpi1}). According to the flavor 
wave function (\ref{DDstarbFlavorWF}) the corresponding isospin amplitudes are
\begin{align}\label{AmplitudeOPE2I}
&\mathcal{M}^{(2)}_{(b1)}=  -4c A \frac{g^4}{f^4} J^g_{22}  \frac{p\cdot\varepsilon p\cdot\varepsilon^*}{p^2-m^2},
\\
&\mathcal{M}^{(2)}_{(b2)}=  4c (d-3)(d-2) A \frac{g^4}{f^4} J^g_{22} \frac{p\cdot\varepsilon p\cdot\varepsilon^*}{p^2-m^2} ,
\\
&\mathcal{M}^{(2)}_{(b3)}=  -4c A \frac{g^4}{f^4} J^g_{22} \frac{p\cdot\varepsilon p\cdot\varepsilon^*}{p^2-m^2} ,
\\
&\mathcal{M}^{(2)}_{(b4)}=  4c (d-3)(d-2) A \frac{g^4}{f^4} J^g_{22} \frac{p\cdot\varepsilon p\cdot\varepsilon^*}{p^2-m^2} ,
\\
&\mathcal{M}^{(2)}_{(b5)}=  4c A \frac{g^2}{f^4} J^c_{0} \frac{p\cdot\varepsilon p\cdot\varepsilon^*}{p^2-m^2} ,
\\
&\mathcal{M}^{(2)}_{(b6)}=  4c A \frac{g^2}{f^4} J^c_{0} \frac{p\cdot\varepsilon p\cdot\varepsilon^*}{p^2-m^2} ,
\\
&\mathcal{M}^{(2)}_{(b7)}=  4 c A \frac{g^2}{f^2} \bigg[\frac{2}{3f^2}\Big(2m^2L+\frac{2m^2}{16\pi^2}\textnormal{log}(\frac{m}{\mu})\Big) \bigg] 
\frac{p\cdot\varepsilon p\cdot\varepsilon^*}{p^2-m^2} ,
\\
&\mathcal{M}^{(2)}_{(b8)}=0,
\\
&\mathcal{M}^{(2)}_{(b9)}=0,
\\
&\mathcal{M}^{(2)}_{(b10+11)}= -\frac32 c A \frac{g^4}{f^4} \big[ (d-2)\partial_\omega J^b_{22}(\omega_1) +\partial_\omega J^b_{22}(\omega_2) \big]  
\nonumber\\
&\qquad\qquad\quad \times \frac{p\cdot\varepsilon p\cdot\varepsilon^*}{p^2-m^2},
\\
&\mathcal{M}^{(2)}_{(b12)}= -\frac32 c A \frac{g^4}{f^4} (d-1)\partial_\omega J^b_{22} \frac{p\cdot\varepsilon p\cdot\varepsilon^*}{p^2-m^2},
\\
&\mathcal{M}^{(2)}_{f^{(2)}}= 4 c A \frac{g^2}{f^2} \bigg[\frac{2}{f^2}\Big(2m^2L+\frac{2m^2}{16\pi^2}\textnormal{log}(\frac{m}{\mu})\Big) \bigg] 
\frac{p\cdot\varepsilon p\cdot\varepsilon^*}{p^2-m^2}, \label{AmplitudeOPE2F}
\end{align}
where $A$ is a constant depending on each diagram and isospin $I$, and we collect them in Table \ref{TabOPEA}. 
Note that in Eq.~(\ref{AmplitudeOPE2F}), $\mathcal{M}_{f^{(2)}}$ stands for the tree-level OPE amplitude 
(Eq.~(\ref{AmplitudeOPE00}) or Eq.~(\ref{AmplitudeOPE01})) where the pion decay constant $f$ has been replaced by $O(\epsilon^2)$ correction $f^{(2)}$.

\renewcommand{\arraystretch}{1.3}
\begin{table}[!htbp]
	\centering
	\caption{The constants $A$ (as well as $\omega_{1,2}$) appearing in the OPE amplitudes (Eqs.~(\ref{AmplitudeOPE2I})-(\ref{AmplitudeOPE2F}). 
	}\label{TabOPEA}
	\begin{tabular}{ccccccccccccc}	\toprule[1pt]
		& $A_{b1}$ & $A_{b2}$ & $A_{b3}$ &$A_{b4}$ & $A_{b5}$ & $A_{b6}$ & $A_{b7}$ & $A_{b8}$ & $A_{b9}$ & $A_{b10+11}$ &$A_{b12}$ & $A_{f^{(2)}}$ \\	\midrule[1pt]
		$I=0$ & $-\frac{3}{16}$ & $-\frac{3}{16}$ & $-\frac{3}{16}$ & $-\frac{3}{16}$ & $-\frac14$ & $-\frac14$ & $\frac34 $ & $-$ & $-$ & $\frac34$ & $\frac34$ & $\frac34$ \\
		$I=1$ & $\frac{1}{16}$  & $\frac{1}{16}$  & $\frac{1}{16}$  & $\frac{1}{16}$  & $\frac{1}{12}$ & $\frac{1}{12}$ & $-\frac14$ & $-$ & $-$ & $-\frac14$ & $-\frac14$ & $-\frac14$ \\
		$\omega_1$ & $\delta$ & $0$ & $-\delta$ & $-\delta$ & $-$ & $-$ & $-$ & $-$ & $-$ & $0$ & $-\delta$ & $-$  \\
		$\omega_2$ & $-\delta$ & $-\delta$ & $\delta$ & $0$ & $-$ & $-$ & $-$ & $-$ & $-$ & $\delta$ & $-$ & $-$  \\
		\bottomrule[1pt]
	\end{tabular}
\end{table}

Final pieces are the TPE contributions depicted in Fig.\ref{O22piDiagram}. Here we need $O(\epsilon^2)$ $DD^*\pi$, $DD^*2\pi$, 
$\bar{D}\bar{D}^*\pi$ and $\bar{D}\bar{D}^*2\pi$ vertexes. Using the chiral Lagrangians (\ref{LagrangianHpi1}) and (\ref{LagrangianHcpi1}),
the isospin amplitudes of TPE are written by
\begin{align}\label{AmplitudeTPE2I}
&\mathcal{M}^{(2)}_{(c1)}= -4 \frac{1}{f^4} \Big[  q_0^2 A_{5} J^F_0 -q_0^2 \big( A_{15}+A_{51}-2A_{5}\big)J^F_{11} + q^2_0 \big(A_{1}
\nonumber\\
&\qquad\quad\ -A_{15}-A_{51}+A_{5} \big)J^F_{21} + \big( A_{1}-A_{15}-A_{51}+A_{5}\big)
\nonumber\\
&\qquad\quad\ \times J^F_{22} \Big]\varepsilon \cdot \varepsilon^*,
\\
&\mathcal{M}^{(2)}_{(c2)}= 4i \frac{g^2}{f^4} \Big[ -q_0 A_5 J^S_{21} + q_0\big(A_1-A_5\big)J^S_{31} + \big(A_1-A_5\big)J^S_{34} \Big]
\nonumber\\
&\qquad\quad\ \times \varepsilon \cdot \varepsilon^* + 4i\frac{g^2}{f^4} \Big[ -q_0 A_5 J^S_{11} + q_0\big( A_1-2A_5\big)J^S_{22} + \big(A_1
\nonumber\\
&\qquad\quad\ -A_5\big)J^S_{24} + q_0\big(A_1-A_5\big)J^S_{32} + \big(A_1-A_5\big)J^S_{33} \Big] 
\nonumber\\
&\qquad\quad\ \times q\cdot\varepsilon q\cdot\varepsilon^*,
\\
&\mathcal{M}^{(2)}_{(c3)}= -4i(d-3)\frac{g^2}{f^4}\Big[ -q_0 \vec{q}^2 A_5 J^S_{11} + (d-2)q_0A_5J^S_{21} +q_0 \vec{q}^2
\nonumber\\
&\qquad\quad\ \times (A_1-2A_5)J^S_{22} + \vec{q}^2 (A_1-A_5)J^S_{24} -(d-2)q_0(A_1
\nonumber\\
&\qquad\quad\ -A_5)J^S_{31} + q_0 \vec{q}^2(A_1-A_5)J^S_{32} + \vec{q}^2(A_1-A_5)J^S_{33} 
\nonumber\\
&\qquad\quad\ - (d-2)(A_1-A_5)J^S_{34} \Big] \varepsilon \cdot \varepsilon^* -4i(d-3)\frac{g^2}{f^4} 
\nonumber\\
&\qquad\quad\ \times \Big[ -q_0A_5J^S_{11} + q_0(A_1-2A_5)J^S_{22} + (A_1-A_5)J^S_{24}
\nonumber\\
&\qquad\quad\ + q_0(A_1-A_5)J^S_{32} +(A_1-A_5)J^S_{33} \Big] q\cdot\varepsilon q\cdot\varepsilon^*,
\\
&\mathcal{M}^{(2)}_{(c4)}=-4i\frac{g^2}{f^4} \Big[ -q_0 \vec{q}^2A_5J^T_{11} + (d-1)q_0A_5J^T_{21} + q_0\vec{q}^2(A_1
\nonumber\\
&\qquad\quad\ -2A_5)J^T_{22} +\vec{q}^2(A_1-A_5)J^T_{24} -(d-1)q_0(A_1-A_5)J^T_{31}
\nonumber\\
&\qquad\quad\ + q_0\vec{q}^2(A_1-A_5)J^T_{32} + \vec{q}^2 (A_1-A_5) J^T_{33} -(d-1)(A_1
\nonumber\\
&\qquad\quad\ -A_5)J^T_{34} \Big] \varepsilon \cdot \varepsilon^*,
\\
&\mathcal{M}^{(2)}_{(c5)}= -4 A_1 \frac{g^4}{f^4} \Big[ \vec{q}^2J^B_{31} - (d+1)J^B_{41} +\vec{q}^2J^B_{42} \Big] \varepsilon \cdot \varepsilon^* + 4A\frac{g^4}{f^4}  
\nonumber\\
&\qquad\quad\ \Big[ J^B_{21} -\vec{q}^2J^B_{22} + (d+3)J^B_{31} - 2\vec{q}^2J^B_{32} + (d+3) J^B_{42}  
\nonumber\\
&\qquad\quad\  - \vec{q}^2 J^B_{43} \Big] q\cdot\varepsilon q\cdot\varepsilon^*,
\\
&\mathcal{M}^{(2)}_{(c6)}=4A_1(d-3) \frac{g^4}{f^4} \Big[ -\vec{q}^2J^B_{21} + \vec{q}^4J^B_{22} -(2d+1)\vec{q}^2J^B_{31}  
\nonumber\\
&\qquad\quad\ + 2\vec{q}^4J^B_{32} + (d-2)(d+1)J^B_{41} - (2d+1)\vec{q}^2J^B_{42} 
\nonumber\\
&\qquad\quad\ + \vec{q}^4J^B_{43} \Big] \varepsilon \cdot \varepsilon^*  - 4A(d-3)\frac{g^4}{f^4}  \Big[ J^B_{21} -\vec{q}^2 J^B_{22} +(d+3) 
\nonumber\\
&\qquad\quad\ \times J^B_{31} -2\vec{q}^2J^B_{32} + (d+3)J^B_{42} - \vec{q}^2J^B_{43} \Big] q\cdot\varepsilon q\cdot\varepsilon^*,
\\
&\mathcal{M}^{(2)}_{(c7)}= -4cA_1(d-3)\frac{g^4}{f^4} J^B_{21} \Big( \vec{p}^2 \varepsilon \cdot \varepsilon^* + p\cdot\varepsilon p\cdot\varepsilon^* \Big),
\\
&\mathcal{M}^{(2)}_{(c8)}= -4A_1\frac{g^4}{f^4} \Big[ \vec{q}^2 J^R_{31} -(d+1)J^R_{41} + \vec{q}^2J^R_{42} \Big] \varepsilon \cdot \varepsilon^* + 4A\frac{g^4}{f^4} 
\nonumber\\
&\qquad\quad\ \times \Big[ J^R_{21} -\vec{q}^2J^R_{22} + (d+3)J^R_{31} -2 \vec{q}^2J^R_{32} + (d+3)J^R_{42} 
\nonumber\\
&\qquad\quad\  - \vec{q}^2J^R_{43} \Big] q\cdot\varepsilon q\cdot\varepsilon^*,
\\
&\mathcal{M}^{(2)}_{(c9)}= 4A_1(d-3)\frac{g^4}{f^4} \Big[ -\vec{q}^2J^R_{21} + \vec{q}^4J^R_{22} - (2d+1)\vec{q}^2J^R_{31} 
\nonumber\\
&\qquad\quad\ + 2\vec{q}^4J^R_{32} + (d-2)(d+1)J^R_{41} - (2d+1)\vec{q}^2J^R_{42} 
\nonumber\\
&\qquad\quad\ + \vec{q}^4J^R_{43}\Big]  \varepsilon \cdot \varepsilon^* - 4A(d-3)\frac{g^4}{f^4} \Big[ J^R_{21} -\vec{q}^2J^R_{22} + (d+3)
\nonumber\\
&\qquad\quad\ \times J^R_{31} -2\vec{q}^2J^R_{32} + (d+3)J^R_{42} -\vec{q}^2J^R_{43} \Big] q\cdot\varepsilon q\cdot\varepsilon^*,
\\
&\mathcal{M}^{(2)}_{(c10)}= 4cA_1 (d-3)\frac{g^4}{f^4} J^R_{21} \Big( \vec{p}^2 \varepsilon \cdot \varepsilon^* + p\cdot\varepsilon p\cdot\varepsilon^* \Big),
\label{AmplitudeTPE2F}
\end{align}
where $A_1$, $A_5$, $A_{15}$ and $A_{51}$ are constants collected in Table \ref{TabTPEA}.

Notice that in above amplitudes (\ref{AmplitudeContact2I})-(\ref{AmplitudeTPE2F}), loop functions $J^{a/b}_{ij}(m,\omega)$, 
$J^{g/h}_{ij}(m,\omega_1,\omega_2)$, $J^{F}_{ij}(m_1,m_2,q)$, $J^{T/S}_{ij}(m_1,m_2,\omega,q)$ and $J^{R/B}_{ij}(m_1,m_2,\omega_1,\omega_2,q)$
are abbreviated as $J^{a/b}_{ij}$, $J^{g/h}_{ij}$, $J^{F}_{ij}$, $J^{T/S}_{ij}$ and $J^{R/B}_{ij}$, respectively. Respective constants
$\omega_1$ and $\omega_2$ for different diagrams are listed in Tables \ref{TabContactA}-\ref{TabTPEA}.
These loop functions $J$ are calculated using dimensional regularization, with the modified minimal subtraction scheme. 

Besides one loop diagrams in Figs.~\ref{O2ContactDiagram}-\ref{O22piDiagram}, at this order $O(\epsilon^2)$ tree-level
amplitudes also emerge. For example, there are $O(\epsilon^2)$ contact contributions that come from $O(\epsilon^2)$ contact Lagrangians
(\ref{Lagrangian4H2h})-(\ref{Lagrangian4H24}). The LECs in Eqs.~(\ref{Lagrangian4H2h})-(\ref{Lagrangian4H24}) actually
consist of two parts: The finite parts that will intrduce large amounts of unknown parameters, and the
divergent parts which are used to renormalize the $O(\epsilon^2)$ 1-loop diagrams. 
In this work, the finite parts of the LECs in Eq.~(\ref{Lagrangian4H2h})-(\ref{Lagrangian4H24}) are ignored due to the lack of fitting data available.

\renewcommand{\arraystretch}{1.5}
\begin{table}[!htbp]
	\centering
	\caption{The constants appearing in the TPE amplitudes (Eqs.~(\ref{AmplitudeTPE2I})-(\ref{AmplitudeTPE2F})). Note that we have $A_{51}=A_{15}$.
	}\label{TabTPEA}
	\begin{tabular}{c|ccc|ccc|cc}	\toprule[1pt]
		& \multicolumn{3}{c|}{ $I=0$  } &\multicolumn{3}{c|}{ $I=1$ } &       & \\
		&  $A_1$   &  $A_5$   &  $A_{15}$   &     $A_1$   &  $A_5$   &  $A_{15}$   &    $\omega_1$   & $\omega_2$ \\
		\midrule[1pt]
		$A_{c1}$ & $\frac{3}{16} $  & $\frac{3}{16} $ & $\frac{-3}{16}$ & $\frac{-1}{16}$ & $\frac{-1}{16}$  & $\frac{1}{16}$ & $-$       & $-$ \\
		$A_{c2}$ & $\frac{-3i}{8}$  & $\frac{3i}{8}$ & $-$              & $\frac{i}{8}$ & $\frac{-i}{8}$  & $-$ & $\delta$  &  $-$ \\
		$A_{c3}$ & $\frac{-3i}{8}$  & $\frac{3i}{8}$ & $-$              & $\frac{i}{8}$ & $\frac{-i}{8}$  & $-$ &  $0$      & $-$ \\
		$A_{c4}$ & $\frac{3i}{8}$   & $\frac{-3i}{8}$ & $-$             & $\frac{-i}{8}$ & $\frac{i}{8}$  & $-$ & $-\delta$ & $-$ \\
		$A_{c5}$ & $\frac{9}{16}$   & $-$             & $-$             & $\frac{1}{16}$ & $-$            & $-$ & $-\delta$ & $\delta$ \\
		$A_{c6}$ & $\frac{9}{16}$   & $-$             & $-$             & $\frac{1}{16}$ & $-$            & $-$ & $-\delta$ & $0$ \\
		$A_{c7}$ & $\frac{9}{16}$   & $-$             & $-$             & $\frac{1}{16}$ & $-$            & $-$ & $-\delta$ & $0$\\
		$A_{c8}$ & $\frac{-3}{16}$  & $-$             & $-$             & $\frac{5}{16}$ & $-$            & $-$ & $-\delta$ & $\delta$ \\
		$A_{c9}$ & $\frac{-3}{16}$  & $-$             & $-$             & $\frac{5}{16}$ & $-$            & $-$ & $-\delta$ & $0$ \\
		$A_{c10}$& $\frac{-3}{16}$  & $-$             & $-$             & $\frac{5}{16}$ & $-$            & $-$ & $-\delta$ & $-\delta$ \\
		\bottomrule[1pt]
	\end{tabular}
\end{table}

In our work, we consider $S$-wave interactions, therefore we have following substitutions for the terms related to the polarization vectors:
\begin{eqnarray}
&&\vec{\varepsilon}\cdot\vec{\varepsilon}^* \rightarrowtail 1, \\
&& \vec{\varepsilon}\cdot \vec{p} \; \vec{\varepsilon}^* \cdot \vec{p} \rightarrowtail \frac{1}{d-1} \vec{p}^2,
\end{eqnarray}

After calculating the scattering amplitudes of $D\bar{D}^*$ systems in four different channels, the $D\bar{D}^*$ potentials in
momentum space can be obtained
via relation:
\begin{align}
\mathcal{V}=-\frac{\mathcal{M}}{\sqrt{\prod 2M_i \prod 2M_f}}.
\end{align}
Note that in  Eq.~(\ref{AmplitudeContact2I})-(\ref{AmplitudeTPE2F}), we did not show the factor $\prod M_i \prod M_f$.

In this work, we aim to investigate weather the $D\bar{D}^*$ interactions are strong enough to form molecular states, therefore we are
interested in $D\bar{D}^*$ potentials at coordinate space. With the help of Fourier transformation we can get potentials $\mathcal{V}(\mathbf{r})$:  
\begin{eqnarray}\label{FourierTransform}
\mathcal{V}(\mathbf{r})=\int \frac{d\mathbf{p}}{(2\pi)^3} \mathcal{V}(\mathbf{p})e^{i\mathbf{p} \cdot\mathbf{r}}.
\end{eqnarray}
Because $\mathcal{V}(\mathbf{p})$ are polynomial in $p$, the integral will be highly divergent with the increasing of the order. Here we renormalize the potential
by introducing a Gaussian cutoff exp$(-\vec{p}^{2n}/\Lambda^{2n})$ with $n=2$ as in Ref.~\cite{Xu:2017tsr}.

\section{Numerical results and Discussions} \label{SecN}

In this section, we first discuss the parameters and the LECs appearing in our calculations. Then we substitute calculated potentials into the Schr\"{o}dinger equation and
explore weather the $D\bar{D}^*$ interactions are strong enough to form bound states.  Later we discuss calculated $D\bar{D}^*$ potentials $\mathcal{V}(\mathbf{r})$
in detail.

In this work, we use the following parameters: $m_\pi=0.139$ GeV, $D$-$D^*$ mass splitting $\delta=0.142$ GeV, 
the decay constant $f_\pi=0.086$ GeV, the renormalization scale $\mu=4\pi f$ and the bare coupling $g=0.65$ as in Ref.~\cite{Xu:2017tsr}.
In our paper we ignore the isospin breaking in the isospin doublet $(D^{(*)0}, D^{(*)+})$. 

For the LECs $D_a$, $D_b$, $E_a$ and $E_b$ in Eq.~(\ref{Lagrangian4H0}), we lack available data that can be fitted, so we
use the resonance saturation to estimate the LECs. With the expressions in appendix \ref{App-1}, we obtain: $D_a = -13.23$, $E_a = -11.49$, $D_b=0$ and $E_b=0$.

\subsection{Bound state properties of $D\bar{D}^*$  } \label{secBindingEnergy}

We first investigate weather $D\bar{D}^*$ interactions are strong enough to form bound states.
the $D\bar{D}^*$ system has the quantum numbers $C=\pm1$ and $I=0,1$, i.e., there are four channels:
$I^G(J^{PC})=0^+(1^{++})$, $0^-(1^{+-})$, $1^-(1^{++})$ and $1^+(1^{+-})$ channels.

For the reliability of our investigation, we will leave the cutoff parameter $\Lambda$ undetermined. Generally, in nucleon-nucleon ChEFT,
$\Lambda$ is adopted below $\rho$ meson mass \cite{Epelbaum:2014efa}. Also, in our previous work, we used $\Lambda=0.7$ GeV \cite{Xu:2017tsr}.
In this paper, we will vary the cutoff at vicinity of $0.7$ GeV to search for the bound state solutions, where a relatively wide range will be adopted.

We solve the Schr\"odinger equations with the $D\bar{D}^*$ 
potentials $\mathcal{V}(\mathbf{r})$ (\ref{FourierTransform}). We present the $\Lambda$ dependences of calculated masses,
the binding energies and the root-mean-square (RMS) radii in Table \ref{TabBoundStateSolutions}.
In Table \ref{TabBoundStateSolutions}, with $\Lambda$ ranging from $0.3$ to $1.1$ GeV, we can see that there are various bound state solutions appearing in considered
channels:

\renewcommand{\arraystretch}{1.5}
\begin{table*}[!htbp]
	\centering
	\caption{The bound state solutions in the four $D\bar{D}^*$ channels. The cutoff parameter $\Lambda$, calculated mass $M$, binding energy $E$\
		and RMS radius $r_{RMS}$ are in units of GeV, MeV, MeV and fm, respectively.
	}\label{TabBoundStateSolutions}
	\begin{tabular}{c|ccc|ccc|ccc|ccc}	\toprule[1pt]
	$\Lambda$	& \multicolumn{3}{c|}{ $0^+(1^{++})$  } &\multicolumn{3}{c|}{ $0^-(1^{+-})$ } & \multicolumn{3}{c|}{ $1^-(1^{++})$ } & \multicolumn{3}{c}{ $1^+(1^{+-})$ } \\
	GeV	&  $M$   &    $E$  &  $r_{RMS}$   &    $M$   &    $E$  &  $r_{RMS}$ & $M$   &    $E$  &  $r_{RMS}$ & $M$   &    $E$  &  $r_{RMS}$  \\
		\midrule[1pt]
	$0.3$	&  $-$   &    $-$  &  $-$   &    $-$   &    $-$  &  $-$ & $-$   &    $-$  &  $-$ & $-$   &    $-$  &  $-$  \\	
	$0.4$	&  $3.875.7$   &    $0.1$  &  $9.0$   &    $-$   &    $-$  &  $-$ & $-$   &    $-$  &  $-$ & $-$   &    $-$  &  $-$  \\
	$0.5$	&  $3874.4$   &    $1.4$  &  $3.2$   &    $-$   &    $-$  &  $-$ & $-$   &    $-$  &  $-$ & $-$   &    $-$  &  $-$  \\
	$0.6$	&  $3872.2$   &    $3.6$  &  $2.1$   &    $-$   &    $-$  &  $-$ & $-$   &    $-$  &  $-$ & $-$   &    $-$  &  $-$  \\
	$0.7$	&  $3870.0$   &    $5.8$  &  $1.8$   &    $3875.6$   &    $0.2$  &  $8.0$ & $-$   &    $-$  &  $-$ & $-$   &    $-$  &  $-$  \\
	$0.8$	&  $3868.6$   &    $7.2$  &  $1.6$   &    $3875.5$   &    $0.3$  &  $6.4$ & $-$   &    $-$  &  $-$ & $-$   &    $-$  &  $-$  \\
	$0.9$	&  $3867.7$   &    $8.1$  &  $1.5$   &    $-$   &    $-$  &  $-$ & $-$   &    $-$  &  $-$ & $-$   &    $-$  &  $-$  \\
	$1.0$	&  $-$   &    $-$  &  $-$   &    $-$   &    $-$  &  $-$ & $-$   &    $-$  &  $-$ & $3862.5$   &    $13.3$  &  $1.0$  \\
	$1.1$	&  $-$   &    $-$  &  $-$   &    $-$   &    $-$  &  $-$ & $-$   &    $-$  &  $-$ & $3813.0$   &    $62.8$  &  $0.6$  \\
		\bottomrule[1pt]
	\end{tabular}
\end{table*}

\subsubsection*{$0^+(1^{++})$}
 For the $0^+(1^{++})$ channel, the bound state solution appears at $\Lambda=0.4$ GeV, and becomes deeper with increasing $\Lambda$. But it then disappears
at 1.0 GeV. The binding energy varies from $0.1\sim8.1$ MeV which is within one order of magnitude. As for the RMS radius, we can see that except large radius (9.0 fm)
at $\Lambda=0.4$ GeV, the radius is basically around 2 fm. Therefore, it is interesting to see that, our obtained bound state solution in this channel is
stable against the variation of $\Lambda$. 

 In a word, in our HMChEFT calculations, the $D\bar{D}^{*}$ interaction
 in the $0^+(1^{++})$ channel is strong enough to form a bound state. In this channel we have a loosely bound state with the mass around 3872 MeV, the binding energy $4\sim5$ MeV and
the RMS radius about $2$ fm. This solution just corresponds to $X(3872)$, so our estimates indicate that $X(3872)$ indeed may be treated
as a good candidate of $0^+(1^{++})$ $D\bar{D}^{*}$ molecular state. Or we can say $D\bar{D}^{*}$ interactions is strongly responsible for $X(3872)$.

\subsubsection*{$0^-(1^{+-})$}
We now focus on the $0^-(1^{+-})$ channel. In Table \ref{TabBoundStateSolutions}, there exists a shallow bound state solution within only a narrow range $\Lambda=0.7\sim0.8$ GeV.
The binding energy $E$ is $0.2\sim0.3$ MeV which is quite small. Consequently, the solution has a large radius with around $7$ fm. Similarly, this solution is also stable
against the variation of $\Lambda$. So we conclude that in the $0^-(1^{+-})$ channel, the binding between $D\bar{D}^{*}$ is relatively weaker comparing to the $0^+(1^{++})$ above,
but their interaction is still strong enough to form a shallow bound state.

\subsubsection*{$1^-(1^{++})$}
In this channel, there is no bound state solutions in the all range ($\Lambda=0.3\sim1.1$ GeV). Combining with the bound state property in the channel $1^+(1^{+-})$ discussed below,
we conclude that the $D\bar{D}^{*}$ interaction in $1^-(1^{++})$ is weakest among all four channels.

\subsubsection*{$1^+(1^{+-})$}
 For the $1^+(1^{+-})$ channel, we can find a bound state solution beginning with a relatively large cutoff $\Lambda=1.0$ GeV. Different from the solutions in previous
 channels, in this channel, the binding energy just becomes deeper and deeper with increasing $\Lambda$. At $\Lambda=1.0$ GeV, we have the binding energy $E=13.3$ MeV with the RMS radius
 $r_{RMS}=1.0$ fm. Because of the large cutoff needed to produce the bound state, we can say that the binding in the $1^+(1^{+-})$ channel is weaker than that in the $0^+(1^{++})$ and
 $0^-(1^{+-})$ channels.
\\
\\
In general, we find that strongest binding of the $D\bar{D}^{*}$ system is in the $0^+(1^{++})$ channel, the next is in the $0^-(1^{+-})$ channel, then in the $1^+(1^{+-})$ channel, the final is
in the $1^-(1^{++})$.

\subsection{The behaviors of the $D\bar{D}^*$ potentials }
In previous subsections, we explore the bound state properties in the four $D\bar{D}^*$ channels. We find that with some reasonable cutoffs, there will emerge
bound state solutions. Some of them are even stable against the variation of $\Lambda$. In this section, to give a deep understandings of the bound state properties
and the $D\bar{D}^*$ interactions, we further discuss the behaviors of the $D\bar{D}^*$ potentials $\mathcal{V}(\mathbf{r})$.

\begin{figure*}[htpb]
	\begin{center}
		\includegraphics[scale=0.5]{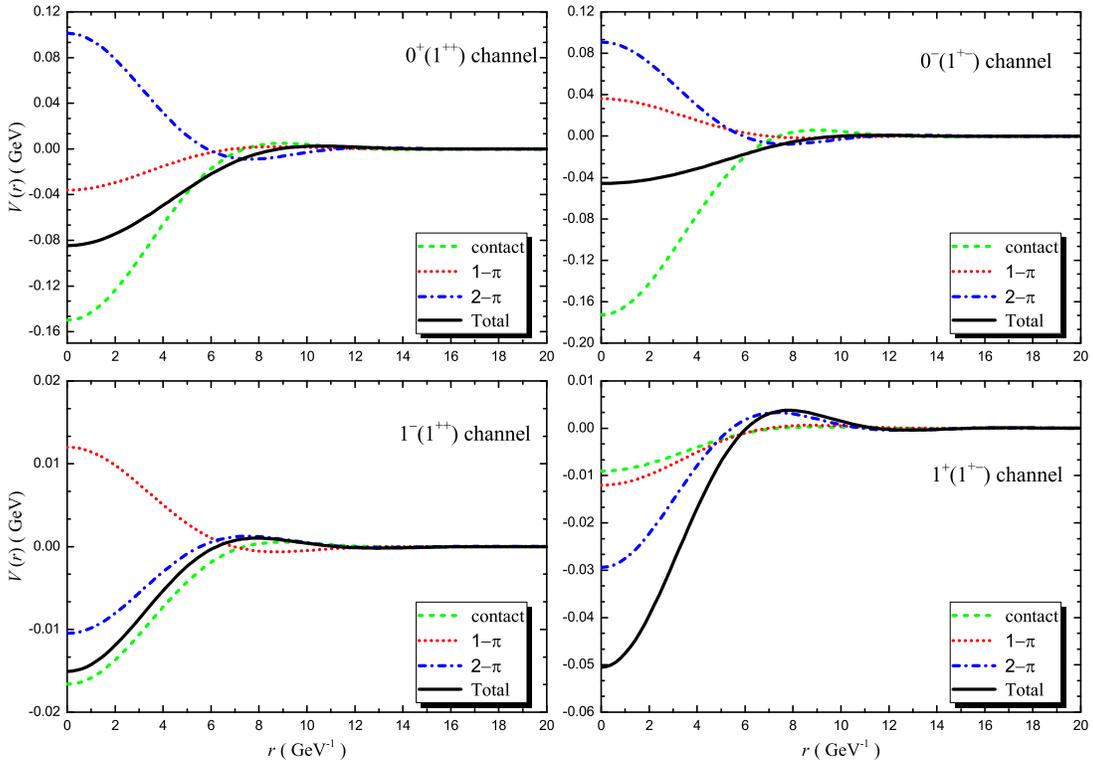}
		\caption{The potentials $\mathcal{V}(\mathbf{r})$ of the $D\bar{D}^*$ system at $\Lambda=0.62$ GeV. Here the $0^+(1^{++})$ channel has a
			bound state solution with mass around $3872$ MeV.
			}\label{FigPotentialVr}
	\end{center}
\end{figure*}

Back to Table~\ref{TabBoundStateSolutions}, we see around $\Lambda=0.6$ GeV the solution in the $0^+(1^{++})$ channel is close to $X(3872)$ most. For a more delicate
investigation, here we set the cutoff to be $\Lambda=0.62$ GeV after fitting the mass of $X(3872)$. With this cutoff we now present $\mathcal{V}(\mathbf{r})$ of all channels
in Fig.~\ref{FigPotentialVr}.

For the $0^+(1^{++})$ channel, we see that the OPE and contact contributions are attractive, where the former is relatively weak. While the TPE contribution provides a
strong repulsive force. Our numerical estimate reveals that with the increase of the cutoff $\Lambda$, the attractive contributions as well as the repulsive 
TPE contribution all become stronger, however their cancellation just leads to a stable total potental $\mathcal{V}(\mathbf{r})$. That is why calculated
binding energy in Table \ref{TabBoundStateSolutions} is stable. 

In the $0^-(1^{+-})$ channel, the behaviors of the contact and TPE contributions do not differ from those in the $0^+(1^{++})$ channel above, whereas the OPE contribution changes to be repulsive.
So, they lead to a weaker attraction comparing to the $0^+(1^{++})$ channel. That explains why the binding energy in this channel is always smaller than that in the $0^+(1^{++})$ channel
(see the binding energies in Table \ref{TabBoundStateSolutions}).

For the $1^-(1^{++})$ channel, the OPE contribution is repulsive while the contact and TPE contributions are attractive. The cancellation between them makes the
line shape of the total $\mathcal{V}(\mathbf{r})$ basically follows the line shape of the contact potential. Observing the amount of the total $\mathcal{V}(\mathbf{r})$, 
we see that although attractive, it is not strong enough comparing to the amount in $0^+(1^{++})$ or $0^-(1^{+-})$ channel.

At the $1^+(1^{+-})$ channel in Fig.~\ref{FigPotentialVr}, all the contributions are attractive. However at the middle range $r=6\sim12$ GeV$^{-1}$, the total $\mathcal{V}(\mathbf{r})$ is
repulsive, which originates from the repulsion of the TPE. This causes the attraction of the $1^+(1^{+-})$ channel to be weakened. Indeed, there is no bound state solution at $\Lambda=0.62$.
Also, from Table \ref{TabBoundStateSolutions} we see that a considerably high $\Lambda$ is needed to produce a bound state solution.

In general, through analyzing the potentials $\mathcal{V}(\mathbf{r})$ depicted in Fig.~\ref{FigPotentialVr}, we have understood the specific mechanism of the bound state formation
in each channel. Depend on
the distinct behaviors of the contact, OPE and TPE contributions in the four channels, the total $\mathcal{V}(\mathbf{r})$s all present attractive line shapes but with different strengths.
We find that most attractive potential is in the $0^+(1^{++})$ channel, then in the $0^-(1^{+-})$, then in the $1^+(1^{+-})$, the final is in the  $1^-(1^{++})$ channel.

Combining with the discussions of the biding energies in Sec.~\ref{secBindingEnergy}, we conclude that there is a hierarchy of the strengths in the $D\bar{D}^*$ interactions:
\begin{align}
\textnormal{str.}[0^+(1^{++})] > \textnormal{str.}[0^-(1^{+-})] > \textnormal{str.}[1^+(1^{+-})] > \textnormal{str.}[1^-(1^{++})],
\end{align} 
where str. stands for the strength of the $D\bar{D}^*$ interaction. Note that despite discrepancies in details, this conclusion is consistent with the one-boson-exchange model calculations
\cite{Liu:2008tn,Zhao:2014gqa,Wang:2021aql}.

\subsection{$0^-(1^{+-})$ and $1^+(1^{+-})$ molecular states}
In previous two sections, we point out the strength ordering in the $D\bar{D}^*$ interactions. But that does not means they can all produce bound states. From Table \ref{TabBoundStateSolutions}
we already fond that only the $0^+(1^{++})$, $0^-(1^{+-})$ and $1^+(1^{+-})$ channels have solutions. In this section we focus on the $0^-(1^{+-})$ and $1^+(1^{+-})$ channels which 
have no direct experimental indications.
 
As discussed in Sec.~\ref{secBindingEnergy}, the solution with mass around $3872$ MeV in the $0^+(1^{++})$ channel just corresponds to $X(3872)$. That means $X(3872)$ has a strong relation
to the $D\bar{D}^*$ interaction, or, $X(3872)$ is a good candidate of the $0^+(1^{++})$ molecular state. In other two channels $0^-(1^{+-})$ and $1^+(1^{+-})$, there also tends to form
molecular states. 

\begin{figure}[htpb]
	\begin{center}
		\includegraphics[scale=0.31]{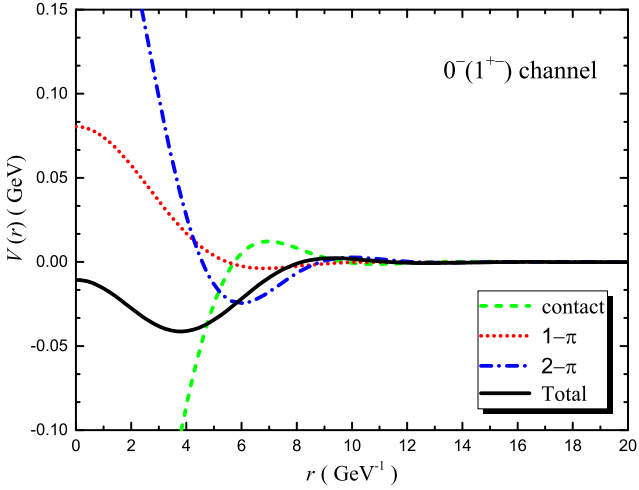}
		\caption{The potentials $\mathcal{V}(\mathbf{r})$ of the $D\bar{D}^*$ system in the $0^-(1^{+-})$ channel at $\Lambda=0.8$ GeV. Here this channel has a
			bound state solution.
		  }\label{FigPotentialVr0m08}
	\end{center}
\end{figure}

From Sec.~\ref{secBindingEnergy}, we have learned that in the $0^-(1^{+-})$ channel, the shallow bound state solution has the binding energy $0.2\sim0.3$ MeV with $\Lambda=0.7\sim0.8$ GeV. Hence we
plot the potential $\mathcal{V}(\mathbf{r})$ at $\Lambda=0.8$ GeV in Fig.~\ref{FigPotentialVr0m08}. From Fig.~\ref{FigPotentialVr0m08}, the OPE contribution is repulsive, therefore the
binding of calculated $0^-(1^{+-})$ molecular state is provided by the contact potential after the cancellation between the contact and TPE contributions. Also, we can see that
the attraction is mainly provided in a range round $2\sim8$ GeV$^{-1}$.

\begin{figure}[htpb]
	\begin{center}
		\includegraphics[scale=0.31]{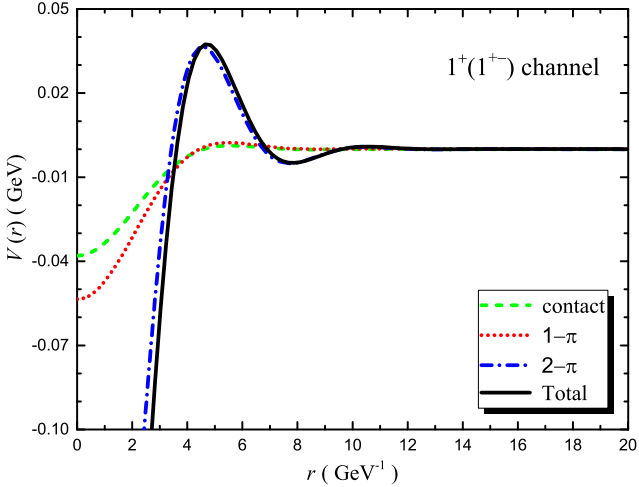}
		\caption{The potentials $\mathcal{V}(\mathbf{r})$ of the $D\bar{D}^*$ system in the $1^+(1^{+-})$ channel at $\Lambda=1.0$ GeV. Here this channel has a
			bound state solution.
		}\label{FigPotentialVr1m10}
	\end{center}
\end{figure}

In other channel $1^+(1^{+-})$, we have a molecular state with considerable binding energy at $\Lambda=1.0$ GeV. Its potential $\mathcal{V}(\mathbf{r})$ is depicted in Fig.~\ref{FigPotentialVr1m10}.
All contributions are attractive, but the binding is mainly provided by the TPE contribution. Also, it is a basically short range attraction. 

In general, the $0^-(1^{+-})$ and $1^+(1^{+-})$ molecular states have different binding mechanisms under the competitions of various contact, OPE and TPE contributions.
As we know, experiment has not observe any possible structures that can fit into our predicted $0^-(1^{+-})$ and $1^+(1^{+-})$ molecular candidates yet, so we briefly discuss their discovery potentials.

We first focus on their decay patterns. The $0^-(1^{+-})$ state has possible two body hidden charm decay channels $\eta_c \omega$ and $J/\psi \eta$. It can also have three body strong decays such as $J/\psi \pi^0 \pi^0$ and $h_c(1P)\pi\pi$.
In other possible channels such as $D^0\bar{D}^{*0}$, $J/\psi \pi^+ \pi^-$ and $D^0\bar{D}^0\pi^0$, the signal can be overwhelmed by $X(3872)$. It may also have
baryonic decay channel $p\bar{p}$, radiative decay channels $\gamma \eta_c$, $\gamma \chi_{c0}$ $\gamma \chi_{c1}$, $\gamma \chi_{c2}$ and $\gamma \eta_c(2S)$.

The $1^+(1^{+-})$ state has possible two body hidden charm decay channels $\eta_c \rho$, $J/\psi \pi$, $h_c(1P)\pi$ and $\psi(2S) \pi$, three body decay channels
 $\eta_c \pi \pi$, $\chi_{c0} \pi \pi$, $\chi_{c1} \pi \pi$, $\chi_{c2} \pi \pi$, baryonic decay channel $p\bar{p}$, and
radiative decay channels $\gamma \eta_c$, $\gamma \chi_{c0}$ $\gamma \chi_{c1}$, $\gamma \chi_{c2}$ and $\gamma \eta_c(2S)$.             
It has more decay channels than the $0^-(1^{+-})$ state, so its decay width may be larger.

These two possible molecular states can be produced in various production mechanisms. For example, they can be searched through subsequent decays in the $e^+e^-$
collision at BESIII and BelleII experiments. It is also promising to collect them in $b$ decay processes such as $B\to X(0^-(1^{+-})/1^+(1^{+-})) +K$.

In fact, some decay channels listed above have already been measured in experiment. For example, $J/\psi \pi$ is the discovery channel of famous
$Z_c(3900)$ \cite{Ablikim:2013mio,Liu:2013dau,BESIII:2015cld,Abazov:2018cyu}. It is possible that our $1^+(1^{+-})$ state is the neutral component of
isovector $Z_c(3900)$, although there exists discrepancy in mass. On the other hand, if the $1^+(1^{+-})$ state is different from $Z_c(3900)$, there may exist more than one enhancements
around $3.8\sim3.9$ GeV in the $J/\psi \pi$ invariant mass spectrum. However because of the limited resolution in the invariant mass distributions of Refs.~\cite{Ablikim:2013mio,Liu:2013dau,BESIII:2015cld,Abazov:2018cyu,Belle:2014nuw,LHCb:2019maw}
we can not pin down this issue now. We hope more sophisticated studies in future can clarify this problem. As for the $h_c(1P)\pi$ and $\psi(2S) \pi$ channel, 
no structures are found around $3.8\sim3.9$ GeV so far.

\begin{figure}[htpb]
	\begin{center}
		\vspace*{-1em}
		\hspace*{-3em}
		\includegraphics[scale=0.41]{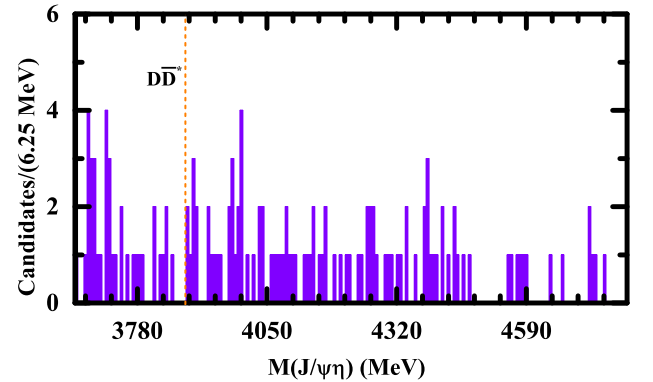}
		\caption{The $J/\psi \eta$ invariant mass distributions in the $B\to J/\psi \eta K$, where the data comes from the $BABAR$ measurement \cite{BaBar:2004iez}.
		}\label{FigBABARJpsieta}
	\end{center}
\end{figure}

The $J/\psi \eta$, which the $0^-(1^{+-})$ molecular state decays into, has also been studied in experiment. In 2004, $BABAR$ measured the invariant mass spectrum of the $J/\psi \eta$ in the process $B\to J/\psi \eta K$ \cite{BaBar:2004iez}.
In the mass distribution of Fig.~\ref{FigBABARJpsieta}, we can see that there seems to have a peak around $D\bar{D}^{*}$ threshold. 
In the measurement of Ref.~\cite{Belle:2013vio}, a small enhancement also can be seen in their $J/\psi \eta$ invariant mass spectrum.  These enhancements may
relate to our predicted $0^-(1^{+-})$ molecular state. A weak sign appears in the $\eta_c \omega$ invariant mass spectrum in Ref.~\cite{Belle:2015yoa} too.
However, in these experiments the events around this region are limited, we hope experiments can pay more attention on these channels, especially the $J/\psi \eta$ in the future.

In addition, $BABAR$ measured kaon momentum spectrum in $B\to X(c\bar{c}) +K$ \cite{BaBar:2019hzd}. $X(3872)$ signal is located around $3.8\sim3.9$ GeV.
With more refined data in the future, we hope there will emerge fine structures if our predicted $0^-(1^{+-})$ and $1^+(1^{+-})$ states exist.

Other production mechanisms such as the low energy $p\bar{p}$ collision at $\overline{\textnormal{P}}$ANDA are also promising. 
The $0^-(1^{+-})$ and $1^+(1^{+-})$ states can be produced through the $s$ or $t$ channel in the $p\bar{p}$ scattering.

\section{Summary} \label{SecS}
In our previous paper \cite{Xu:2017tsr}, we studied the double charmed system $DD^*$ in the framework of heavy meson chiral effective field theory (HMChEFT). We
applied obtained $DD^*$ chiral interactions to exotic state investigations. Predicted molecular state is consistent with the latest finding $T_{cc}$ \cite{LHCb:2021vvq}.
In present work, we extend it to investigate the charmed-anticharmed system $D\bar{D}^{*}$. 
As mentioned in Sec.~\ref{sec1}, various $XYZ$ states were observed around open-charm thresholds, therefore the elaborate heavy-meson interactions studied here
may help us to reveal the formations of these charmonium-like states.

The generalization of previous work is not straightforward. So we first construct $D\bar{D}^{*}$ interaction Lagrangians by properly considering charge conjugation properties
of the fields in the heavy quark limit. We then calculate $S$-wave $D\bar{D}^{*}$ potentials up to $O(\epsilon^2)$ order
at 1-loop level using weinberg's scheme. The complete contact, one pion exchange (OPE) and two pion exchange (TPE) interactions are included. For a deeper understanding of the
$D\bar{D}^{*}$ interactions, we further solve the Schr\"odinger equations using calculated potentials. 

The behaviors of the $D\bar{D}^{*}$ potentials as well as bound state properties are analyzed in detail. In considered four channels i.e. $I^G(J^{PC})=0^+(1^{++})$, $0^-(1^{+-})$, $1^-(1^{++})$ and $1^+(1^{+-})$,
there appears distinct behaviors depending on the competitions of the contact, OPE and TPE interactions. This leads to the specific mechanism of the bound state formation
in each channel. For example, in the $0^+(1^{++})$ and $0^-(1^{+-})$ channels, relatively large repulsions of the TPE contributions cause them to have weak bindings.

Combining the potential behaviors and the bound state properties, we conclude that there exists a strength ordering of the $D\bar{D}^{*}$ interactions in the four channels which would lead
to different binding abilities:
\begin{align}
\textnormal{str.}[0^+(1^{++})] > \textnormal{str.}[0^-(1^{+-})] > \textnormal{str.}[1^+(1^{+-})] > \textnormal{str.}[1^-(1^{++})],
\end{align}
where str. stands for the strength of the interaction.

Further, our investigation reveals that the $D\bar{D}^{*}$ interaction in the $0^+(1^{++})$ channel is strongly responsible for $X(3872)$ in experiment, in specific,
 $X(3872)$ can be a good candidate of the $0^+(1^{++})$ molecular state. In addition, there also tends to form molecular states in the $0^-(1^{+-})$ and $1^+(1^{+-})$ channels.

We examine the potential behaviors of the $0^-(1^{+-})$ and $1^+(1^{+-})$ molecular states in detail, point out their formation mechanisms. Their discovery potentials
are also discussed. We hope future experiments such as BESIII, LHCb and BelleII can further study them in the $e^+e^-$ productions and $b$-decays,
to search for the predicted multi-structures around the $D\bar{D}^*$ mass region. We also expect that they can be searched in future $\overline{\textnormal{P}}$ANDA experiment.

\section*{Acknowledgments}
We thank Fu-Lai Wang, Zhan-Wei Liu and Xiang Liu for helpful discussions. 
This work is supported by the National Natural Science Foundation of China (NSFC) under Grant No. 12005168.

\appendix

\section{Determination of the unknown LECs}\label{App-1}

Assuming contact contributions from Eq.~(\ref{Lagrangian4H0}) are equivalent to
$\rho$, $\phi$ and $\sigma$ exchanges, we will be able to extract $D_a$, $D_b$, $E_a$ and $E_b$.

For $\rho$, $\phi$ and $\sigma$ exchanges, we need corresponding interacting Lagrangians for $D^{(*)}D^{(*)}V$ vertex:
\begin{align}\label{LagrangianHV}
	&\mathcal L_{HV}=
	i\beta\langle H v_\mu (-V^\mu) \bar H \rangle
	+i\lambda \langle H \sigma_{\mu\nu} F^{\mu\nu}(\rho) \bar H\rangle,
	\\
	&\mathcal{L}_{H\sigma}=g_s \langle H\sigma \bar H\rangle , \label{LagrangianHS}
\end{align}
as well as their charge conjugations:
\begin{align}
	&\mathcal{L}_{HcV}= -i\beta\langle \bar{H}_c  v_\mu
	(-V^\mu) {H}_c\rangle
	+i\lambda\langle \bar{H}_c
	\sigma_{\mu\nu}F^{\mu\nu}(V)H_c\rangle, \label{LagrangianHcV}
	\\
	&\mathcal{L}_{Hc\sigma}=g_s \langle \bar{H}_c\sigma H_c\rangle, \label{LagrangianHcS}
\end{align}
where $F_{\mu\nu}(V)=\partial_\mu V _\nu - \partial_\nu V_\mu +
[V_\mu,{\ } V_\nu]$ and
\begin{eqnarray}
	V&=&\frac{ig_V}{\sqrt2}\left(\begin{array}{cc}
		\frac{\rho^{0}}{\sqrt{2}}+\frac{\omega}{\sqrt{2}}&\rho^{+}\\
		\rho^{-}&-\frac{\rho^{0}}{\sqrt{2}}+\frac{\omega}{\sqrt{2}}
	\end{array}\right).
\end{eqnarray}
In above, $H$ and $H_c$ has been defined in Eq.~(\ref{Hfield}) and (\ref{Hcfield}), $g_V=5.8$, $\beta=0.9$ and $\lambda=0.56$ GeV$^{-1}$
\cite{Li:2012ss},
$g_s=\frac{3.73}{2\sqrt{6}}$ \cite{Bardeen:2003kt}.

By matching the $D\bar{D}^*\to D\bar{D}^*$ amplitudes in the two ways, we get
\begin{align}\label{LECcontact}
	&D_a\sim-\frac{g_s^2}{m_\sigma^2}-\frac{\beta^2g_v^2}{4m_\omega^2}, \quad E_a\sim-\frac{\beta^2g_v^2}{4m_\rho^2},\quad D_b\sim 0, \quad E_b\sim 0.
\end{align}

\end{document}